\documentclass[11pt]{amsart}

 \usepackage[margin=1in]{geometry}

\usepackage[numbers]{natbib}
\usepackage{url}
\usepackage{comment}

\usepackage{amssymb, amsmath, amscd, amsthm, color, epsfig, url, tikz, graphicx
}

\usepackage[T1]{fontenc}

\usepackage{caption}
\usepackage[all]{xy}          
\xyoption{dvips}              

\makeatletter
\renewcommand\l@subsection{\@tocline{2}{0pt}{2pc}{5pc}{}}
\makeatother

\theoremstyle{plain}

\theoremstyle{definition}

\newtheorem{def/ex}[thm]{Definition/Example}

\theoremstyle{remark}

\makeatletter
\@namedef{subjclassname@2020}{%
  \textup{2020} Mathematics Subject Classification}
\makeatother

\begin{document}


\title{The New Mathematics of Democracy}


\author{Bailey Flanigan}
\address{Department of Political Science and Computer Science (LIDS), MIT,  Cambridge, MA 02139}
\email{baileyf@mit.edu}
\urladdr{baileyflanigan.com}

\author{Ismar Voli\'c}
\address{Department of Mathematics \& Statistics and Institute for Mathematics and Democracy, Wellesley College, Wellesley, MA 02481}
\email{ivolic@wellesley.edu}
\urladdr{ivolic.wellesley.edu}

\subjclass[2020]{Primary 91B14; Secondary 91B12, 91B32, 62P25, 90C27, 68T07}

\keywords{voting theory, computational social choice, proportional representation, participatory budgeting, equal shares, deliberative democracy, sortition,  artificial intelligence}


\maketitle

\begin{abstract}

This article surveys emerging directions in the mathematics of democracy. It uses three case studies --- voting theory, participatory budgeting, and deliberative democracy --- to highlight how contemporary challenges motivate rigorous mathematical research that incorporates real-world data, institutional constraints, and implementation feasibility. Within each case, we highlight active and promising research frontiers, evidence of real-world impact, practical applications, and opportunities for getting involved.
\end{abstract}

\section{Introduction}

Democracy has always presented mathematical questions: processes like tallying votes, apportioning seats, and deciding district lines are fundamentally mathematical. For centuries, mathematicians have studied these mechanisms, developing rich and celebrated theories of voting, representation, fairness, power, and collective choice. 

Our democratic landscape has changed considerably over this period. We now live in an environment shaped by intense polarization, declining trust in institutions, fragmented media ecosystems, politically driven litigation, compartmentalized information flow, and erosion of longstanding democratic norms. At the same time, new systems of democratic participation like ranked choice voting, proportional representation, citizens’ assemblies, participatory budgeting, and AI-assisted democratic tools are moving from theory into practice, often by popular demand from disillusioned citizens. 

The shifting contexts and venues in which we practice democracy present the need and the opportunity to innovate on the mathematical foundations of democracy. As we will illustrate in this article, the “messiness” of practical real-world democracy introduces new mathematical questions and pushes the required theory and methodology into new directions beyond classical domains. We need models that are more grounded in political behavior; theoretical methods that can integrate real-world data, uncertainty, institutional context, and AI tools; and theories that engage with how democratic mechanisms are implemented, used, undermined, and perceived.

As we will illustrate with research already being done, these needs lead to new open questions that are mathematically rich and diverse, requiring ideas from across probability, statistics, optimization, algorithms, graph theory, dynamical systems, geometry, analysis, and elsewhere. Of course, mathematical design is only part of the challenge; even the most principled mathematical interventions must respond to the realities of the institutions within which they embed, and the populations which they serve. For this reason, the mathematical work we describe is best done in partnership with scholars across the social sciences and law, the practitioners who implement these technologies, and the public officials who will interface with citizens through them.

This article illustrates these points through three practically impactful case studies: \textbf{voting}, \textbf{participatory budgeting}, and \textbf{deliberative democracy}. This is not an exhaustive survey; mathematics also plays important roles in districting and gerrymandering, legislative apportionment, election auditing and security, the measurement of voting and legislative power, polling and forecasting, and many other aspects of democratic systems.\footnote{Districting deserves special mention. Over the last decade, mathematical work on districting has been exceptionally successful and lauded, creating an entire new branch of applied mathematics. Mathematicians, computer scientists, political scientists, and legal scholars have produced exciting results and methods for assessing and thinking about gerrymandering, often in conjunction with real-world election, census, and other demographic data. They have partnered with legislatures and democracy organizations, and have curated public-facing districting tools like Districtr \citep{Districtr}. This vibrant body of work has already been surveyed in many accessible expositions aimed at broad audiences \citep{Veomett24, Hershberger22, Orcutt23, NYTDuchin25}.}$^,$\footnote{There are many more democratic applications of mathematics across computational social choice and mechanism design, including choosing representative committees (\textit{committee selection}) and dividing resources fairly (\textit{fair division}). These fields now encompass a literature far too large to review here. For an introduction, see \citep{HandbookCOMSOC} as well as the COMSOC community website \citep{COMSOC} for information on current research, conferences, courses taught, and publications.}  We focus on these three cases as areas where research is actively developing, there is a demonstrated possibility of real-world impact, and opportunities for contributions are especially rich. Each case study traces the same arc, illustrating how classical theory has addressed practical democratic challenges and highlighting open questions offering further research opportunities.

\section{Voting theory}

Voting theory  is arguably the most developed democratic application in which mathematicians have made crucial contributions. Since the 1770s, when French mathematicians Jean-Charles de Borda and Marquis de Condorcet proposed two methods for tallying ranked ballots (although both had been described before their time), voting theory has expanded into a rich field that now occupies a central place in economics, political science, and mathematics. 

In \emph{ranked choice} (or \emph{preferential}) voting systems, voters rank candidates in order of preference rather than selecting a single favorite, as they do in \emph{plurality} voting. The collection of preferences, or a \emph{profile}, is then subject to some procedure that selects the winner.  Borda suggested assigning points to the rankings, with first place on each ballot earning $k-1$ points, where $k$ is the number of candidates, each second place getting $k-2$ points, etc. The candidate with the most points across the profile is the winner. The Condorcet method considers how many voters placed candidate $A$ over candidate $B$, or the other way around, for each pair of candidates $(A,B)$. If a candidate wins all of their head-to-head matchups, they are the winner.\footnote{This is really a family of methods; since pairwise matchups can produce cycles of preferences, various \emph{Condorcet completion} methods, or ways to break out of the cycle if one occurs and choose the winner, have been proposed over the years.}

A third method, \emph{instant runoff voting} (IRV), evolved as a special case of the \emph{single-transferable vote} (STV) system described by English lawyer Thomas Hare in the 1850s. The latter is designed for selecting multiple winners while the former is its single-winner variant. Unlike Borda and Condorcet, these are currently used in political elections in the U.S.~and elsewhere around the world.  IRV is an iterative algorithm: if no candidate wins a majority of first-place votes, the candidate with the fewest first-place votes is eliminated and their votes are transferred to the second choices of voters who had the eliminated candidate at the top of their ballots. This retabulation might now yield a majority winner, but if it does not, the process is repeated.\footnote{There are many other methods of tallying ranked ballots, but these three and their hybrids are the best known and most studied.}

To understand how these voting rules compare, consider the following example. Suppose there is an election with eleven voters and four candidates $A$, $B$, $C$, and $D$. One voter expresses the preference $A>C>B>D$, one $A>C>D>B$, four $C>A>B>D$, and five $B>D>A>C$. Then $A$ is the Condorcet winner (they beat $B$ 6-5, $C$ 7-4,  and $D$ 6-5), $B$ is the Borda winner (they have 20 points, $A$ has 19, $C$ has 16, $D$ has 11), and $C$ is the IRV winner (eliminate $D$, then $A$, and $C$ finally beats $B$ 6-5).

In the mid-twentieth century, voting theory became part of the broader area of 
\emph{social choice theory}, the formal study of how individual preferences aggregate into group decisions, originating from the intersection of welfare economics and political economy. Social choice theory axiomatizes voting, laying out some desirable criteria that a preferential voting system might satisfy. Given a finite set of \emph{alternatives} (or \emph{outcomes}, or \emph{candidates}) $A$ with $|A|\geq 3$, a set $[n]=\{1,2,...,n\}$ of \emph{agents} (or \emph{voters}), and the set $\mathcal W(A)$ of all (weak) orderings $R$ on $A$, social choice theory studies the properties of \emph{social welfare functions}
$$
F\colon \mathcal W(A)^n\longrightarrow \mathcal W(A).
$$
An element of $\mathcal W(A)^n$ is understood as a ranked choice profile and a social welfare function is a voting method producing a ranking of the candidates for each profile.\footnote{A modification is to set the target of $F$ to be the set of non-empty subsets of $A$, in which case one gets a \emph{social choice function} that produces a subset of the winning alternatives rather than a (weak) ranking of all of them.} Borda, Condorcet (completion methods), 
and IRV are all social welfare functions.

The cornerstone of social choice theory is the Impossibility Theorem \citep{Arrow1951} proved by Kenneth Arrow in 1951 (earning him the Nobel Prize in 1972), showing that even basic democratic desiderata can be mutually incompatible. More precisely, the theorem states that no social welfare function can satisfy the following three conditions at once:

\begin{itemize}
\item {\bf Non-dictatorship}: There is no agent $d\in [n]$ such that, for every profile $R$, whenever $d$ strictly prefers one alternative to another, $F(R)$ does as well.

\item {\bf Pareto efficiency}: For every $R$, if every agent ranks alternative $x$ above $y$, so does $F(R)$.

\item {\bf Independence of irrelevant alternatives}: For every pair of alternatives $x$ and $y$, the way $F(R)$ ranks $x$ and $y$ depends only on how voters rank $x$ and $y$. I.e.~if two profiles $R$ and $R'$ have the same relative rankings of $x$ and $y$, then $F(R)$ and $F(R')$ rank $x$ and $y$ in the same way.

\end{itemize}

With Arrow's Theorem, the focus in social choice theory largely shifted from designing systems to understanding their properties and mathematical limitations. Over the last seventy years, a tremendous volume of work produced a wealth of results about the behavior of voting methods. The construction of pathological profiles proliferated,
revealing quirky features of voting methods, failures of desirable criteria, and opportunities
for strategic manipulation by candidates and voters.

But classical social choice theory studied voting methods in a highly idealized setting, since the goal was not necessarily to model actual elections but to formally understand preference aggregation. For instance, classical models assume the completeness of orderings, which in practice means that voters never submit partially filled ranked ballots (including \emph{bullet votes} with only the top choice indicated), and that every profile contains a preference ordering for every voter. 

Arrow's Theorem itself also requires \emph{universal domain}, namely that $F$ be defined on all of $\mathcal W(A)^n$ so that all profiles are possible and should be considered simultaneously. This assumption is brought into question in actual elections where, given the ideological positions of candidates and voters, not all profile configurations are realistic. Another common assumption is that all voters lie on the same one-dimensional ideological spectrum, and their preferences are \emph{single-peaked} --- each voter has an ideal preferential peak and their preferences decrease with distance from that position. Under this assumption, the Median Voter Theorem \citep{B48}, another seminal result in social choice theory, says that the top preference of the median voter beats all other candidates in head-to-head matchups. Since the median voter is often near the political center, this theorem is cited by Condorcet advocates as evidence that this method would encourage candidates to adopt more centrist platforms that reduce polarization and discourage fringe or extreme political posturing \citep{FM25}.

\textbf{From idealized models to real elections.}
The nature of social choice theory's assumptions experienced a shift in the 
late twentieth century, when researchers began increasingly turning to computational methods and randomness. By generating large numbers of synthetic elections under various models, voting theorists could compare voting rules and estimate the frequency of the failure of social choice criteria. Common early examples include spatial models, in which voters and candidates occupy points in a metric space and voters rank candidates according to distance; impartial culture, in which voters independently choose rankings uniformly at random; and impartial anonymous culture, which places a uniform distribution on anonymous preference profiles.

These models shifted consideration from what \textit{could} happen in principle to what is \textit{likely} to happen, but this work still relied on highly stylized assumptions about the distribution of votes. More recent work extended this idea, developing parametrized generative models intended to incorporate more realistic electoral conditions. Plackett-Luce  and Bradley-Terry models, for example, can be incorporated into generative frameworks that allow the tuning of turnout rates, candidate strength, polarization, bloc cohesion, and support across groups.\footnote{Plackett–Luce assigns a ``strength'' to each candidate and  rankings are generated with probability proportional to that strength; Bradley–Terry considers pairwise preferences determined by relative strengths of the candidates.}
The parameters can be estimated from data or chosen in consultation with practitioners, so that these models now serve as a bridge between theory and real-world elections. Collaborations with litigators, reform organizations, and community groups have resulted in case studies and white papers \citep{ElectionSystemCaseStudies} as well as software packages like VoteKit \citep{DDGGHMW25} and Preferential Voting Tools \citep{HP25} which provide infrastructure for generating, importing, and analyzing elections under a range of empirically informed models.

Recent years have also seen a shift toward empirical and computational social choice research anchored in real elections and survey-based models. One direction uses a database of about 3,000 (and growing) IRV and STV elections from the U.S., Scotland, and Australia \cite{BardalBrillMcCunePeters2025,   DeshpandeGargJacobson2026, DickersonMartinMcCune2024, GrahamSquireMcCune2025, GrahamSquireZayatz2021, KatzmanDougherty2026, KilgourGregoireFoley2020, McCuneGrahamSquire2024, McCuneMcCune2024, McCuneNaber2026,Schwab_report, McCuneWilson2026,Stephanopoulos24, TomlinsonUganderKleinberg2023}. These have been used to show, for example, that IRV and Condorcet methods choose the same winner over 99\% of the time  and almost always select a strong candidate. They perform much better across numerous measures than Borda and various other methods, including plurality which unsurprisingly performs the worst. Behaviors like the spoiler effect and vote splitting are extremely rare for both, with IRV marginally underperforming on spoiler compared to Condorcet. Strategic moves like truncating ballots and burying candidates (ranking them low) are non-issues for IRV and present a slight weak point for Condorcet methods, while the effect of ballot length restrictions, which are common in RCV elections, appears to depend on the extent of the restriction and the electoral context.\footnote{A legitimate criticism of using IRV and STV cast-vote data to study other voting methods is that, if those other methods were actually used, voters and candidates might behave differently, producing profiles and yielding results different from the ones at hand.}

\begin{figure}[h]
\centering
\begin{tabular}{cc}
\includegraphics[width=70mm]{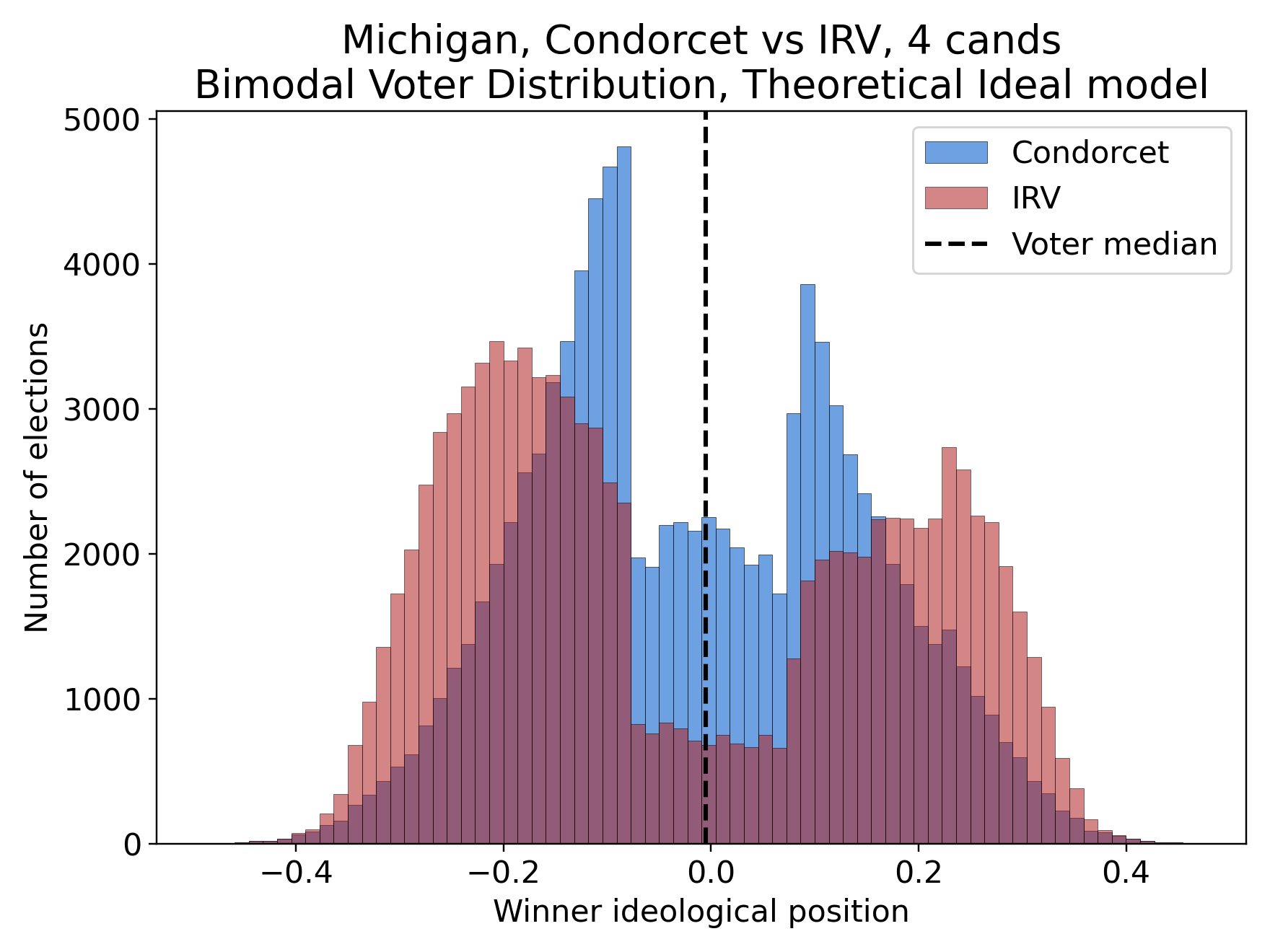} & 
\includegraphics[width=70mm]{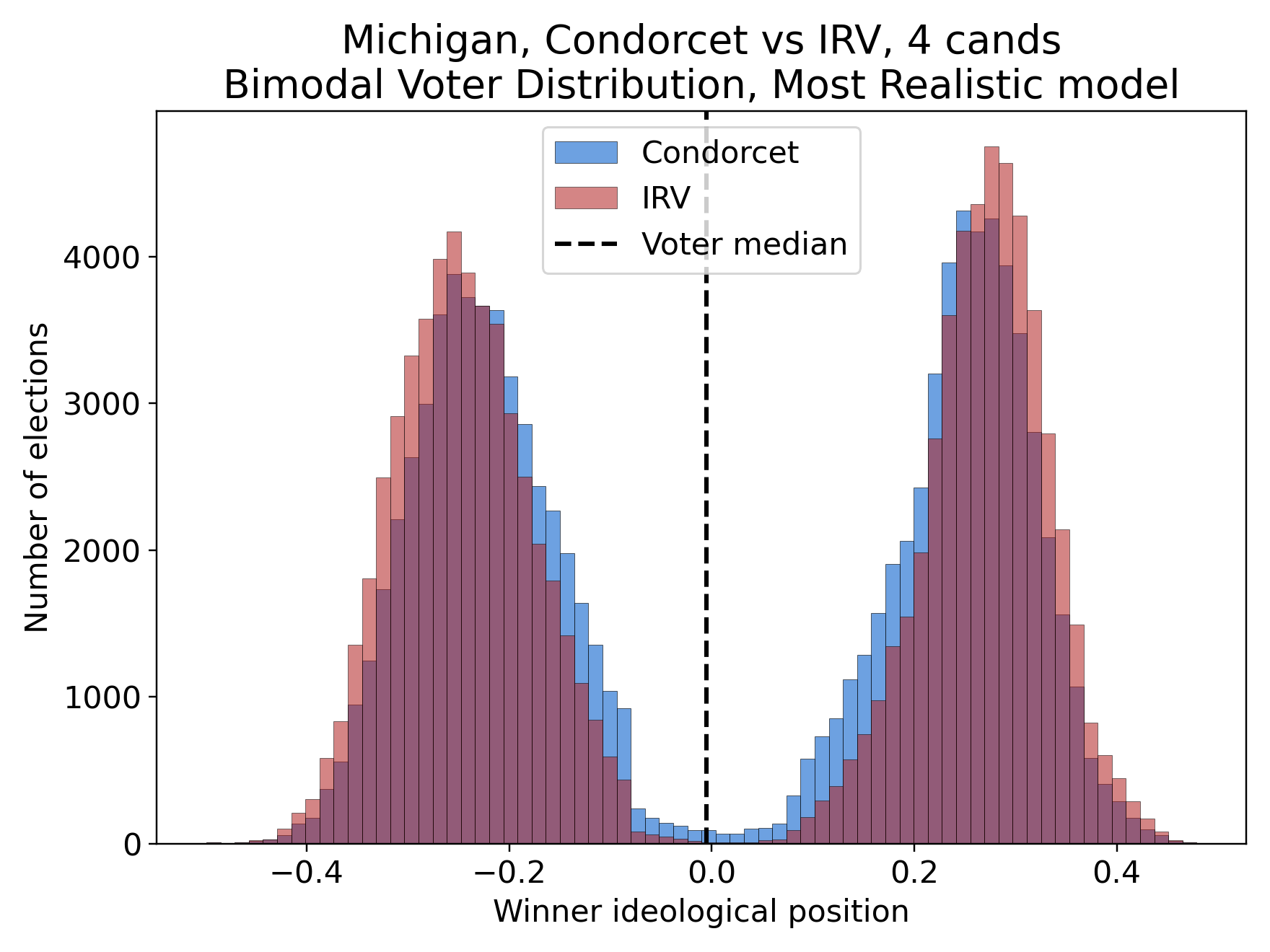}\\
\end{tabular}
\caption{Histograms showing the ideological positions of Condorcet and IRV winners across 100,000 simulated elections with four candidates, based on the CES survey in Michigan. Theoretical Ideal model assumes that all voters cast a ballot, provide complete preferences, and know the exact ideological placement of every candidate. Most Realistic model incorporates truncated ballots, voter abstention, and noise in voters' perception of the candidates' ideological position.}
\label{fig:MI_results_bimodal}
\end{figure}

New research also incorporates synthetic electorates, building them from real elections or large surveys such as the American National Election Studies \citep{ANES} or Cooperative Election Study (CES) \citep{SAS} which contain questions on ideology, party affiliation, and hot-button issues. These questions are used to construct voter distributions and generate elections that mimic actual voter preferences. As an example, one pioneering work  uses CES-based distributions to demonstrate that Condorcet methods elect centrist candidates more often than IRV \citep{AFG}. However, like much of the other literature in this vein  (with notable exceptions such as \citep{HKRW, KilgourGregoireFoley2020}), this work assumes idealized voting conditions including full turnout and complete ballots. Subsequent work \citep{CandidateModeration} imitates voter behavior more closely by, for example, building the actual observed bullet vote rate of about 35\% into the model. With these additional conditions, the difference in the selection of moderate candidates between IRV and Condorcet largely disappears; an illustration can be found in Figure \ref{fig:MI_results_bimodal}. The discrepancy between the conclusions of these two papers illustrates the need for importing real-world simulation into research.

\textbf{Single transferable vote (STV) and proportional representation.} So far, we have focused on single-winner elections, but ranked ballots can be extended to elections with multiple winners using STV. The central mathematical question driving a growing body of research on STV is \emph{proportional representation},  particularly in connection with districting. The main problem in this arena is that, even without the presence of gerrymandering, U.S.~single-winner districts are prone to diluting votes and producing unrepresentative results, especially with plurality elections and the defective system of party primaries.\footnote{About a third of Massachusetts voters, for example, vote Republican, but they are not able to elect a single representative to Congress because of how spread out they are throughout the state \citep{DGHKNW19}.} Moreover, two recent Supreme Court decisions have reshaped the landscape: the 2019 Rucho v.~Common Cause case, which held that partisan gerrymandering claims are nonjusticiable political questions beyond the reach of federal courts, and the 2026 Louisiana v.~Callais ruling, which sharply curtailed the use of Section 2 of the Voting Rights Act to protect minority representation. These paved the way for the ongoing gerrymandering wars and the mid-decade redistricting frenzy, exposing the need for examining alternatives that are structurally shielded from political machinations.

One such alternative is \emph{multi-member districts}, which are larger districts that elect several representatives at once. One way to elect such a slate of candidates is STV, which takes ranked ballots as input and sets a quota that has to be cleared in order to win a seat. This is typically the \emph{Droop quota}, given as $\left\lfloor \frac{n}{k+1}\right\rfloor+1$ where $k$ is the number of seats to be filled and $n$ is the number of voters. When $k=1$, this reduces to the usual simple majority. If a candidate passes the quota with their first-place votes, they are elected. Surplus votes beyond the quota are redistributed to the next-place candidates in a proportional way. If no candidate reaches the quota, IRV-style elimination and transfer of votes occurs. The quota and the surplus redistribution make STV a \emph{proportional} system, meaning that a cohesive group of voters who rank the same slate of candidates above all others and who make up at least $p$ quotas of the electorate can secure the election of about $p$ of those candidates.\footnote{Most democracies use multi-member districts, including at-large ones, to achieve proportional representation, but voters usually express preference for parties rather than candidates. The most widely used of these \emph{party list} systems are D'Hondt and Sainte-Laguë. In the U.S., these are known as Jefferson and Webster methods and were both used in the past to apportion the House of Representatives. A way to translate between the two pairs is to think of states as parties and their populations as votes for that state/party.}

New work on multi-member districts and STV is exceptionally encouraging. Simulations across all 50 states show that this system significantly curtails the possibility of partisan gerrymandering \citep{GGRS21}. Even more remarkably, it produces far more proportional representation for racial and ethnic minorities than single-winner districts, and this proportionality even holds without race-conscious line-drawing; i.e.~race-blind ``neutral'' maps perform as well as maps optimized for minority representation \citep{BenadeBuckDuchinGoldWeighill2021, MGGG_FRA}. An empirical analysis of the 2024 Portland, OR STV city council elections illustrates these potential benefits as people of color were able to elect candidates of choice in every city district, with STV ensuring no single bloc of like-minded voters could sweep the results anywhere \citep{DDL_Portland}.\footnote{One hurdle in implementing multi-member districts is that they were banned for Congressional elections by the 1967 Uniform Congressional District Act. The reason was that bloc plurality voting was used to elect the slates of candidates, and this method allows a unified majority to win every seat. The Congress was concerned that some states would take advantage of this weakness, dissolving their districts and making all their elections at-large so as to deny representation to minorities \cite{Plier2020}. But instead of changing the voting system, the 1967 Act threw out the baby with the bath water, discarding the entire multi-member system when the real culprit was the tallying rule.}

These results motivate broader questions of  how representational quality should be measured. Beyond asking whether a rule is proportional in principle, research now  looks at how faithfully different multiwinner systems translate the support of political, racial, geographic, and other groups into seats. This has brought new tools, including metric geometry, into the study of proportional representation, with analyses of how representational fidelity depends on district magnitude, voter cohesion, electoral geography, and district design \citep{BenadeBuckDuchinGoldWeighill2021,GGRS21,MGGG_FRA}.

\subsection{Frontier work and open questions}

The developments described above illustrate a common trend in contemporary voting theory: the field is moving beyond the analysis of idealized preference profiles toward mathematical models grounded in real election data. Building on these ideas, we now highlight three rich open directions in voting theory.


\textbf{Voter behavior models.} Existing empirical work still captures only a fraction of the behaviors observed in real elections. Voters abstain, truncate  ballots or fill them out incorrectly, respond to campaigns, vote strategically, and react to conflicting or changing information. Candidates adjust their positions, enter or exit races, and form coalitions in response to electoral incentives. Developing mathematical models that incorporate these phenomena remains an important challenge. 
In addition to the approaches described above, one recent approach adapts a \textit{smoothed model} \cite{spielman2004smoothed} to the ranked voting context; here, voting profiles are worst case, but are perturbed by a small amount of neutrally-structured noise. Axiomatic analysis in this model reveals which axiom violations are brittle to noise and which are robust, and has been applied to many classical axiomatic impossibilities, including Arrow's theorem \cite{flanigan2023smoothed, xia2020smoothed,xia2021semirandom,xia2023condorcet}. While the smoothed model goes beyond the worst case while still hedging against problematic structures, the noise model that permits its guarantees is still subject to strong assumptions, and it does not incorporate real-world data. Valuable advances in this area would combine insights from existing models to develop data-driven formal models of voter behavior, and then determine which conclusions about voting rules remain robust under these more realistic models \citep{AFG,COLLW25,DDGGHMW25,CandidateModeration}. 

\textbf{Preference formation and opinion dynamics.} Classical voting theory typically treats voter preferences as fixed inputs to an election. In reality, those preferences emerge through campaigns, media, social networks, deliberation, and interpersonal interaction before ballots are cast. Understanding elections therefore requires understanding how preferences form and evolve. Network and dynamical models have begun to describe polarization, algorithmic recommendation systems, and (mis)information flow \citep{Baumann2020, Peralta2021, Peralta2024}. These models raise questions about polarization, convergence, stability, and the relationship between individual interactions and collective outcomes. Related models of candidate positioning examine when electoral competition draws candidates toward the political center and when it instead produces separated ideological clusters. Graph theory and dynamical systems provide natural tools for studying these evolving networks, while sheaf-theoretic models can represent situations in which people communicate different information in different social contexts \citep{HansenGhrist2021}.  A major challenge is integrating these models with voting rules themselves, allowing researchers to study how evolving preferences translate into democratic outcomes.


\textbf{Voting advice applications.} As voters increasingly have the option to use chatbots for on-demand voting advice, another important area of study is \textit{voter advice applications}: online systems that recommend candidates or parties to voters, typically based on voter and candidate responses to policy questionnaires.\footnote{Examples include StemWijzer (Netherlands), Wahl-O-Mat (Germany), Smartvote (Switzerland), and Vote Compass (Canada, Australia, the United States, and elsewhere).} Recent work has begun giving the voter advice pipeline mathematical treatment. For example, adaptive questionnaires choose which question to ask next as a function of a
voter's previous answers, trading off the amount of information elicited against
the accuracy of the eventual recommendation \cite{Bachmann2024}. Other work considers how these pipelines may be vulnerable to manipulation;
candidates can change how they report their positions, platform designers can choose the metric and weights used to compute similarity, and question designers determine which dimensions of political disagreement enter the calculation at all. Using data
from the Swiss platform Smartvote, one line of investigation identifies several such
vulnerabilities and shows that seemingly small choices at each of these stages can
substantially change the platform's recommendations
\cite{Berdoz2025}. Since a voting advice application is itself an elicitation and aggregation mechanism, it naturally lends itself to social choice-style axiomatic analysis, raising many open questions for future work.

\section{Participatory budgeting}

A newer democratic application where mathematics has been valuable is participatory budgeting
(PB). PB is a form of voting, except that instead of voting over discrete candidates, citizens are
collectively deciding how to spend a fixed amount of public funds. PB originated in Porto Alegre, Brazil in 1989 \cite{de2005participatory}, where it has been linked to gains in sanitation and infant mortality \citep{goncalves2014}. From roughly a hundred Brazilian cities, PB has since spread to more than 7,000 jurisdictions worldwide \citep{pbp_about}, running in communities as small as Greensboro, NC, which allocates about \$100,000 for PB per city district, and in cities as large as New York, Milan, Glasgow, Madrid, Seoul, Toronto, and Paris, which allocated a total of roughly 75 million euros in 2023 \cite{pb_votes_list}.

Implementations of PB vary, but they generally include the following phases: citizens suggest projects they would like to see funded; a smaller group meets to refine the list of projects; and then the broader population votes on the projects to be funded. The primary contributions of mathematics so far have been in voting over projects, so we focus primarily on that
stage here and broaden our view in Section \ref{PBfrontier}.

At the heart of PB voting is a classical combinatorial optimization problem known as the \textit{knapsack problem.} There is a set of $m$ predefined projects, each with a cost defined by a cost function $c:[m]\to\mathbb{R}_{\geq 0}$, where $[m]=\{1,2,\ldots,m\}$. There is a fixed budget $B$, and the goal is to collectively select a \textit{budget-feasible} bundle of projects, i.e., a set of projects $S\subseteq[m]$ satisfying
\[
    \sum_{j\in S} c(j)\leq B.
\]
Thus, the feasible outcomes are exactly the feasible solutions to a knapsack problem; the crucial difference is that an ordinary knapsack problem also specifies a numerical value for each item and wants to maximize the value of the selected set. In PB, there is no such exogenously given objective; instead, there are $n$ voters, each of whom submits a ballot communicating some preferences about which projects to fund. What 
makes one feasible bundle better than another is then based on how well it reflects the preferences of these voters.

This introduces two distinct mathematical problems. First, \textit{how should those preferences be represented and elicited?} In the most general formulation, each voter can have an arbitrary ranking or utility function over feasible project bundles; unfortunately, there may be exponentially many such bundles, making a complete description of these preferences challenging. Second, \textit{once voters' preferences have been represented in some tractable way, how should they be aggregated to choose one feasible bundle?} One might want to maximize total satisfaction, as in the knapsack problem; however, as we will see, it may also be democratically desirable to give guarantees about how benefits are distributed across voters. Much of the mathematical work on PB can be understood as addressing one or both these questions, and we give a practically influential example for each below.

\textbf{A case study on representing preferences: Knapsack voting.}
Most of the PB literature makes the problem of eliciting preferences tractable by imposing simple structure on those preferences. The most common model is the approval model, in which each voter simply approves or disapproves of
each project, while other models equip voters with a strict ranking or cardinal utilities over projects \citep{aziz2021participatory,fairstein2023realworld}.
Approval-based models then translate to \textit{approval ballots} which ask voters to approve as many projects as they want; a $k$-approval ballot instructs voters to approve exactly $k$ projects. Approval-based representations and ballots have the advantage of being simple, but they can discard crucial information related to the budget constraint --- which projects a voter would give up in order to fund others. Indeed, if a voter can approve $k$ projects that cost more than the available budget, we learn little about how the voter would make the necessary trade-offs.

\textit{Knapsack voting}, introduced by Goel et.~al.~in 2015 \cite{goel2015knapsack,goel2019knapsack}, provides a richer ballot format that aims to address this gap. In knapsack voting, each voter submits a budget-feasible bundle $S_i\subseteq[m]$, thereby solving their own version of the knapsack problem. Votes of this form are collected into a profile $S_1,\ldots,S_n$, and are then typically aggregated by scoring each project by the number of voters who included it (sometimes scaled by its cost). Projects are funded in decreasing order of score, skipping any project that would exceed the remaining budget.

The formal justification for knapsack voting is clearest in the\textit{ continuous relaxation} of PB. In this setting, each project is modeled as being divisible into small ``dollar units,'' so a voter's ballot corresponds to how they would spread the budget over these projects continuously. Welfare then has a simple
geometric meaning: considering a voter's ideal budget allocation as a point in space, a voter is better off when the selected budget allocation has smaller distance from their ideal point. Social welfare is the sum of these utilities across voters, so the welfare-maximizing outcome is the allocation that is, in aggregate, closest to voters' ideal budgets. In this fractional model, knapsack voting is strategy-proof
and welfare-maximizing \cite{goel2019knapsack}. For indivisible projects, the feasible region becomes discrete, and these guarantees hold in the discrete case to a strong approximate degree.

Since its introduction, knapsack voting has moved beyond a purely theoretical proposal. Data from the Stanford Participatory Budgeting platform document 32 elections that used knapsack voting as the primary ballot \citep{gelauff2024rank}. 
Similar ``shopping cart'' voting methods have also been used in European PB processes, including in Madrid and Reykjavik
\citep{goel2019knapsack}.  Evaluation of knapsack voting in real-world data reveals behavior consistent with the theory: voters often select cheaper projects under knapsack ballots than under $k$-approval \citep{gelauff2024rank}. Although this evidence suggests that knapsack ballots are addressing a real gap in how voters engage with budget-induced trade-offs, a voter's favorite budget-feasible bundle is still only one point in an exponentially large space of
possible outcomes. It does not tell us a voter's ranking over other feasible bundles, their marginal values for projects, or how they view substitutions and complementarities. Understanding what richer preference
structures can be represented and elicited without asking voters to report an exponentially large object
is therefore a broader mathematical problem, to which we return in Section \ref{PBfrontier}.

\textbf{A case study on aggregating preferences: Method of Equal Shares.} Most real-world PB elections --- including those using approval-style ballots and knapsack ballots --- are tallied by a greedy aggregation rule: projects are simply sorted by their vote totals (or alternatively, votes per cost) and funded in that order until the budget is exhausted \citep{fairstein2023realworld,goel2019knapsack}. Such a rule implicitly gives the collective knapsack problem a simple objective which is to maximize the total project-level support. This is transparent and easy to implement, but it neglects a potentially important democratic principle: ensuring that the funded projects serve all voters to a proportional degree. Indeed, under greedy rules, it is easy to show that certain groups may get a large share of the benefit, while others get nothing.

The Method of Equal Shares (MES), introduced by Peters, Pierczyński, and Skowron in 2021 \cite{peters2021proportional}, seeks to rectify this problem. It still uses standard approval ballots where voter $i$ approves a subset of projects $A_i\subseteq[m]$, but it changes the aggregation mechanism to directly pursue proportionality.\footnote{MES can also be used with general additive utilities, in which case it satisfies an approximation of the EJR guarantee described below. For simplicity, we focus our description here on the standard approvals version.} In MES, the public budget is treated
as if it were first divided equally among voters so that each voter begins with a virtual account of
$B/n$. Starting from $S=\emptyset$, the rule repeatedly looks for a not-yet-funded project whose cost can be paid
for by the voters who support it, using only the virtual money they have left.

More formally, in a given round, for each not-yet-funded project $j$, let $N_j=\{i\in[n]:j\in A_i\}$ be the set of voters who approve $j$, and let $r_i\leq B/n$ be voter $i$'s remaining virtual budget. The rule
computes the smallest price that can be paid per voter who approved project $j$:
\[
    \rho_j
    =
    \inf_{\rho\geq 0}\left\{
        \sum_{i\in N_j}\min\{r_i,\rho\}\geq c(j)
    \right\}.
\]
If no such $\rho$ exists (i.e., the feasible set in the optimization above is empty), then $j$ is not currently affordable. Otherwise, $j$ can be funded by charging each approving voter $i\in N_j$ the amount $\min\{r_i,\rho_j\}$. MES selects an affordable project with the smallest value of $\rho_j$, funds it, and deducts these payments from the approving voters' accounts. If no project is affordable for any $\rho$, the main rule stops, often followed in
practice by a completion step to make sure the budget is used up to the maximum possible degree
\citep{peters2021proportional}.

Intuitively, this procedure seeks to ensure everyone gets their due share of satisfaction from the chosen projects. This is MES's core justification, and it is mathematically formalized by the fact that MES provably satisfies the axiom of \textit{extended justified
representation (EJR)} \citep{peters2021proportional}. Intuitively, EJR ensures that groups with sufficiently cohesive opinions receive an appropriate level of representation in the funded outcome, with the guarantee strengthening as the group grows. EJR can be viewed as a relaxation of a natural stability requirement: a sufficiently
large subgroup should not be able to use its proportional share of the public budget to purchase a different bundle of projects that it collectively prefers. EJR belongs to a larger family of \textit{justified-representation} axioms that are broadly useful in set selection problems, formalizing the informal goal that ``all groups should get their fair share'' in the chosen set.

MES has been used in real participatory budgeting elections in cities across
the Netherlands, Poland, and Switzerland between 2023 and 2025 \citep{equalsharesWebsite,wikipediaMES}. 
In such cases, the proportionality satisfied by MES had demonstrable empirical impacts. In Wieliczka, Poland, relative to the standard greedy approval rule described
above, MES reduced the share of voters receiving none of their approved
projects from \(28\%\) to \(18\%\). It also funded projects in the southern part of the municipality, which the standard greedy rule would have shut out entirely \citep{equalsharesWieliczka2023}. In Aarau, Switzerland, MES selected at least one project from every city district, while the greedy method would have only selected projects that were either city-wide or from the central district
\citep{equalsharesAarau2023}.


This line of work also opens several mathematical questions about what axioms or other desirable goals can be achieved simultaneously with proportionality. MES may unnecessarily leave part of the budget unspent, so practical versions require completion
rules; this has motivated work on efficient and principled completion procedures
\citep{kraiczy2023adaptive,kraiczy2025streamlining}. There are also incentive questions, since proportional rules may incentivize voters to approve only their highest-priority projects, effectively making themselves harder to satisfy. There may be further trade-offs between proportionality and welfare --- the axiom prioritized by greedy rules. 
These are just a few of the many practically-important axiomatic trade-offs in PB (e.g., \cite{papasotiropoulos2025bounded}), leaving open a large field of mathematical questions around how to define and achieve a ``good'' PB solution.

\subsection{Frontier work and open questions}\label{PBfrontier}
An encouraging feature of the PB landscape is that the infrastructure for testing and deploying new voting methods already exists. The Stanford Participatory Budgeting platform, which has supported PB elections across many U.S. municipalities, implements several of the ballot designs discussed here — including k-approval, knapsack, and ranked formats — and releases ballot data that feeds back into empirical research \cite{gelauff2024rank}. Another rich data source for evaluating new methods is \textit{Pabulib}, an open library of standardized datasets from hundreds of real participatory budgeting elections \citep{faliszewski2023participatory}. Given the promise of this application area, we highlight two emerging areas of research.

\textbf{New preference models.} As described above, a major tension in PB lies between the need to ask voters simple questions and the massive space of project bundles that then must be chosen from. The main ballot types are simple, and thus implicitly assume preferences that have simple structure. A major direction of PB innovation is the development of richer preference models, which in turn motivate richer ballot formats and new voting methods for aggregating them. For example, 
%
one line of work aims to model dependencies between projects, in the form of substitutes and complements \citep{goyal2023mechanism} or
broader interaction structures such as project groups, category-level budget limits, and logical constraints such as dependencies or incompatibilities among projects (e.g., \citep{rey2025constraints}). In a way, EJR itself requires project interdependency, except in its current conception, this is an axiom imposed by the election designer rather than something inherent to voters' preferences. Some empirical evidence suggests that voters do prefer more proportional bundles over greedy ones \cite{yang2024designing}, but integrating this axiomatic idea into a formal preference model constitutes rich future work.


\textbf{Project list design.} Another major challenge in PB happens \textit{upstream} of the vote: the process going from a large collection of submitted proposals to the refined
set of $m$ projects over which the vote occurs. This stage dictates the PB agenda and has an immense amount of power over the ultimate election result, yet it is comparatively understudied. This raises a \textit{two-stage} set selection problem: how should projects be proposed, refined, priced, merged, and shortlisted into an initial set, which will then be passed through a voting rule to choose the final set?

A small literature has begun to formalize this broader pipeline. One line of inquiry formally models the shortlisting and voting stages in succession, studying both how to construct a shortlist and the strategic incentives created by the interaction
between the two stages \citep{rey2021shortlisting}. Other work treats project submission itself as a game among strategic proposers \citep{faliszewski2025projectsubmission}, allows proposers to strategically choose project costs \citep{faliszewski2025strategiccost}, or allows an agent to add and remove projects to cause a target project to win \citep{faliszewski2025projectstrength}.

The problem of list design becomes especially salient as computational tools increasingly mediate that process. Machine-learning and natural-language-processing tools have been developed to classify, summarize, and filter
citizen proposals, including in real PB settings
\citep{arana2021citizen,davies2021nlp,shin2026machinelearning,zambrano2025llms}. These tools make it possible to process proposal spaces that would otherwise be unwieldy, but they also raise questions about what formal algorithmic guarantees are desirable, and which ones existing tools satisfy. Examples of such guarantees might deal with what kinds of diversity should be preserved in the shortlisting process, how robust the final allocation is to changes or errors in this procedure,  and which proportionality axioms or other axioms can be achieved in the end-to-end process, rather than in just the final vote.

\section{Deliberative democracy}

The examples above treat democratic decision-making primarily as a problem of
\textit{aggregation}: voters express preferences over candidates, projects, or
bundles, and a voting rule converts those preferences into an outcome.
Deliberative democracy begins from a different premise: many democratic
decisions require citizens to not only register preferences, but to learn,
exchange reasons, weigh trade-offs, and revise their views in light of other
perspectives \citep{dryzek2019crisis,gutmann2004why}. In fact,
deliberation can be used to enrich the types of decisions from the previous
two sections. For example, in ballot-measure elections, the Oregon Citizens'
Initiative Review repeatedly convened a small panel of citizens to study a ballot initiative,
hear from advocates and experts, deliberate, and produce a voter-facing statement unpacking the options \citep{gastil2020hope}.
Participatory budgeting processes also often include deliberative stages, particularly during the process of narrowing down the project set for the final vote \citep{gilman2012transformative}.

More often, deliberation is used to give citizens a say in democratic decisions that
are \textit{not} reducible to a single vote. \textit{Deliberative
mini-publics} --- including citizens' assemblies, citizen juries, consensus
conferences, and deliberative polls \cite{fishkin2000deliberative,goodin2006deliberative,oecd2020} --- bring together a relatively small group
of everyday people, usually selected by lottery and stratified to reflect the wider
public, to learn about a policy issue and develop recommendations.
Over the past few decades, deliberative mini-publics have gained substantial momentum and political influence worldwide. Collectively, the OECD \citep{oecd2020}, Participedia \citep{participedia}, and the POLITICIZE dataset \citep{paulis2021} have documented well over a thousand such processes across the local, regional, state, national, and supranational levels, with the pace of new assemblies surging sharply since 2010. Increasingly, these bodies are moving toward binding influence and permanent institutionalization. In Ireland, a Citizens' Assembly recommended repealing the constitutional ban on abortion, a change ratified by national referendum in 2018
\cite{farrell2021reimagining}. 
In East Belgium, the Ostbelgien Model established the world's first permanent citizens' council with standing agenda-setting power \citep{niessen2022}. 

Although deliberative democracy may initially sound less mathematical than
counting votes, it introduces a wide range of formal questions: how to select a
representative panel of deliberation participants, divide participants into
productive discussion groups, model changes in opinion, systematically explore large spaces of potential proposals, and aggregate incomplete or evolving judgments. The most mathematically mature solutions respond to the first question --- selecting deliberation participants --- so we focus primarily on that here. Afterwards, we turn to several emerging directions.

\textbf{Sortition.} In the typical case, the participants of a deliberative
mini-public, called a \textit{panel}, are selected by \textit{sortition}, i.e., by random
lottery. In its idealized
conception, sortition is simply a \textit{uniform} lottery, giving everyone in the
population an equal chance to occupy one of $k\in\mathbb{N}$ seats on the
panel. In practice, however, sortition is complicated by selection bias:
different population groups decline invitations to participate at different
rates, so a panel selected uniformly from the volunteers would likely skew,
for example, toward older and more highly educated people
\cite{procaccia2022}. Organizers therefore usually (1) invite a large
uniform sample of residents; (2) collect those who respond affirmatively into a
\textit{pool} of volunteers; and (3) randomly draw the final panel from this pool with the requirement that the panel must satisfy demographic quotas reflecting the population's composition.

The most technically interesting step is (3), and it is formalized as follows. Let $N=[n]$ be the pool of volunteers collected in step (2). Represent a panel
$P\subseteq N$ by its incidence vector $x^P\in\{0,1\}^n$, where $x^P_i = 1$ indicates that pool member $i$ is included in the panel $P$. Let
$\mathcal{G}$ be a collection of groups defined by age, gender, geography,
race, education, or other features, with integer lower and upper quotas
$\ell_g$ and $u_g$ for each $g\in\mathcal{G}$. The set of \textit{feasible} panels is then
\[
    \mathcal{P}
    =
    \left\{
        P\subseteq N:
        |P|=k
        \text{ and }
        \ell_g\leq |P\cap g|\leq u_g
        \text{ for every }g\in\mathcal{G}
    \right\}.
\]
Equivalently, the incidence vectors of feasible panels consist of integer coordinates $x_i$ 
satisfying
\[
    \sum_{i=1}^n x_i^P=k,
    \qquad
    \ell_g\leq\sum_{i\in g}x_i^P\leq u_g
    \quad\text{for every }g\in\mathcal{G}.
\]
Importantly, these constraints are typically imposed on \textit{marginal} groups, i.e., those defined by one feature at a time (e.g., ``women'' or ``people from the south''), though generic combinations of features are permitted and sometimes used. Because of these overlapping quotas, the sortition problem is not simply a task of sampling from mutually exclusive strata, and even deciding whether a
feasible panel exists is NP-complete
\cite{flanigan2021fairalgorithms}.

A sortition algorithm chooses not just a feasible panel, but a probability
distribution $D\in\Delta(\mathcal{P})$ over feasible panels. For a panel $P$ randomly drawn from $D$, the first moment of the distribution over the corresponding incidence vector is the vector of individual selection probabilities
\[
    \pi
    =
    \mathbb{E}[x^P],
    \qquad
    \pi_i
    =
    \Pr[i\in P].
\]
As motivated by a long tradition of political theory along with the
requirements of practice, much of the sortition literature aims to make the
coordinates of $\pi$ as equal as possible
\cite{baharav2024goldilocks,flanigan2021fairalgorithms,
flanigan2020neutralizing}. Game-theoretic arguments later showed that controlling
inequalities among these probabilities is also essential to preventing manipulation by pool members via misreporting one's attributes \cite{baharav2024goldilocks, flanigan2024manipulation}, providing an additional mathematical justification for equality.

To consider how equal one can make a given lottery, we must consider the set of attainable first moments, which corresponds to the convex polytope
\[
    \mathcal{Q}
    =
    \operatorname{conv}\{x^P:P\in\mathcal{P}\}
    \subseteq [0,1]^n.
\]
This perspective reveals why one can optimize individual selection probabilities without enumerating the potentially astronomical collection $\mathcal{P}$. By Carath\'eodory's theorem, every point of $\mathcal{Q}$ can be implemented by a lottery supported on only a small number of panels. Equality-optimizing algorithms exploit this geometry through column generation \cite{flanigan2021fairalgorithms} by optimizing over the convex hull of an iteratively constructed collection of panels; dual variables define a linear objective used to search for a new feasible panel, and certify optimality when no improving panel exists.

The first moment, however, does not determine the full distribution. Two lotteries can have the same vector $\pi$ but very different \textit{joint} selection probabilities
\[
    q_{ij}
    =
    \Pr[i,j\in P]
    =
    \mathbb{E}[x_i^P x_j^P].
\]
Leaving these correlations uncontrolled may be democratically undesirable, as small-support lotteries may repeatedly select the same people
together, diminishing the overall randomness in the sortition process. Recent
work instead maximizes the Shannon entropy
\[
    H(D)
    =
    -\sum_{P\in\mathcal{P}}D(P)\log D(P)
\]
of the distribution over panels, possibly subject to constraints on the
individual selection probabilities \citep{deazevedo2026maximallyrandom}. Here,
Carath\'eodory's sparse representation is no longer applicable,
since entropy depends on the full distribution rather than only its first
moment. In this case, convex duality is used to transform the problem into one involving weighted sums over feasible panels, while dynamic programming makes the required counting and sampling tractable
\citep{deazevedo2026maximallyrandom}. 

Extensions address further requirements of practice, using additional mathematical tools. Work on making sortition algorithms \textit{publicly transparent} rounds an optimized lottery $D$ into a uniform distribution over a smaller, publicized collection of panels. This work uses discrepancy theory to bound the changes in selection probabilities resulting from this rounding, thereby bounding the change in the lottery's optimality \citep{flanigan2021transparent}. Additional work on \textit{alternates} jointly selects panelists and replacements to fill in 
representation gaps left when participants drop out after being selected. This work uses empirical risk minimization and learning-theoretic bounds to obtain a near-optimal solution from sampled dropout scenarios \citep{assos2025alternates}.

In summary, a single, practice-driven question --- how to fairly draw and
maintain a panel --- has required tools from convex geometry, optimization,
game theory, dynamic programming, discrepancy theory, and learning theory.
This breadth of mathematics is matched by an unusually direct path from
theory to implementation, given deliberative democracy practitioners' strong enthusiasm for adopting new technical methods. The original equality-optimizing algorithm was implemented in \textit{StratifySelect}, the Sortition Foundation's open-source
selection software, and has been used to recruit many real citizens' assemblies \citep{sortitionFoundation2021fairest}. Many of these tools --- and more --- are also hosted on \texttt{Panelot.org}, a free, open-source platform hosted by the algorithm developers. Over the course of 2024 and 2025, Panelot was used on
294 distinct days to draw 1{,}071 assemblies.

\subsection{Frontier work and open questions} 


Within and beyond sortition, deliberative democracy offers many mathematical questions and applications. A new direction in sortition considers how to achieve richer representation notions than those reflected by categorical demographic quotas, modeling participants' attributes in a continuous metric space
\citep{ebadian2022sortition,ebadian2025boosting}. Another set of works considers how to divide participants into discussion groups that balance various combinatorial criteria \citep{barrett2023talking}; mathematically, this becomes a constrained submodular-maximization problem over repeated
partitions.
Other work designs interaction protocols based on formal models of how local interactions shift viewpoints, optimizing for
the quality of the resulting outcome relative to an underlying social optimum
\citep{fain2017sequential,goel2025metric,munagala2025matching}.

Beyond these procedural design questions, one important set of complex open questions in deliberative democracy arises from the fundamental fact that the space of policy options is massive. This makes deliberation categorically different from ordinary voting problems. The alternatives are not fixed in advance, but are generated and refined during the process itself. A mechanism must therefore explore the space of possible proposals while simultaneously eliciting and aggregating preferences over them.

Recent work illustrates the role for mathematics --- often interleaved with machine learning tools --- in helping build such mechanisms. For example, one line of work on \textit{generative social choice} combines justified-representation guarantees with language models to generate a set of statements that proportionally represent participants, based on participants' free text inputs \citep{fish2024gsc}. A related tool, called the \textit{Habermas Machine,} uses language
models to synthesize statements that participants with different views may
jointly endorse \citep{tessler2024habermas}. Another class of platforms, including \textit{Pol.is} \citep{small2021polis} and \textit{Remesh} \citep{masoodalavi2022peacebuilding}, approach the same problem via \textit{decentralized}
participant input; through an online platform, citizens generate short position statements and
evaluate statements proposed by others, so both the columns and observed entries of the resulting participant--statement matrix arise from the same
electorate. The goal is to take this highly sparse preference information and, through methods like clustering, sampling, and optimization, map the opinion space and find comments with high cross-group support--known as \textit{bridging} \cite{blair2026structure}. The key technical challenge is that because each participant evaluates only a small fraction of all
statements, this type of partial preference data requires careful technical handling of a sparse approval matrix  
\citep{halpern2024computing,halpern2023representation}. While these methods already use adaptive queries, further work could tailor adaptive querying to richer downstream representation goals.

These tools are just the beginning; they leave many opportunities for new algorithms and axiomatic guarantees, as well as unexplored assumptions about the structures of participants' opinions --- how they manage incomplete or conflicting preferences, how they learn and persuade, and how they vote with incomplete access to the space of options. Mathematics can be valuable not only in building tools to handle more complex opinion structures, but also in formalizing such tools' implicit assumptions and their consequences (as in, e.g., \cite{flanigan2026pluralism}).

\section{Conclusions}\label{AI}

We hope the case studies in this article convey two sets of ideas. First,
real democracy can be both a consumer and a source of mathematics. The mathematics we have described here arose because an idealized model was brought into conversation with an actual democratic context. Second, deciding what a mechanism
\emph{should} do remains a political and normative question, and 
mathematics is thus most valuable when used in conjunction with insights and real-time input from political scientists,
legal scholars, practitioners, and the citizens whose participation these
mechanisms are meant to serve.  For mathematicians who want a way in, the infrastructure is welcoming. Across these applications, there are open datasets and online platforms where new solutions are being deployed and new questions are arising.

%
While there is great anxiety about how AI will change both research and democracy, we argue that the growing use of AI makes mathematics even more important to the democratic endeavor.
This might seem counterintuitive, because much of democratic participation is naturally expressed in language: citizens explain what they desire, describe trade-offs, object to proposals, suggest compromises, and articulate values that do not fit neatly into a ballot. Language models make it newly possible to process this kind of input at scale, and this creates a real opportunity for more participatory democratic institutions. However, this also creates a temptation to collapse the entire problem into the model --- collect free-text input, ask an LLM to summarize it, and treat the result as democratic judgment. 

That would be a mistake, amounting to outsourcing democratic judgment to AI rather than using AI to deepen \textit{people's} democratic judgment. Achieving the latter is where mathematics becomes essential. Formal models, assumptions, axioms, and guarantees provide a scaffold within which AI can operate while its role remains explicit and controlled. Systems that integrate AI and math in this way (e.g., as proposed in \citep{flanigan2026pluralism}) can use AI not as the democratic decision-maker, but as a \textit{translator} between the richness of human expression and mathematical objects, which can then be used to facilitate new democratic methods while preserving human judgment and public inspection. 

The difficulty of mathematics for democracy is that elegant theory must meet what it means, in practice, for a system to be truly ``democratic.'' That difficulty is also the opportunity: when mathematics takes the features of real-world democracy seriously, it can help expand the range of democratic institutions we know how to design and implement. At a time when democracy must respond to rapidly shifting challenges, the field is open, the questions are urgent, and their consequences extend far beyond mathematics.

{\small
\bibliographystyle{plain}
\bibliography{bibliography}
}

\end{document}